\documentclass[11pt,twoside]{article}
\usepackage{asp2010}
\usepackage{natbib}

\resetcounters

\begin{document}

\title{Interstellar dust}
\author{Compi\`egne M.$^1$
\affil{$^1$Canadian Institute for Theoretical Astrophysics, University of Toronto, 
             60 St. George Street, Toronto, ON M5S 3H8, Canada}
}

\begin{abstract}
Dust is a key component of the Universe, especially regarding galaxies evolution,
playing an essential role for both the physics and chemistry of the interstellar medium. 
In this paper, we give a brief review of interstellar dust. 
We describe the main dust observables and how 
it allows us to constrain dust properties.
We discuss the dust lifecycle and the dust evolution in the ISM.
We also present a physical dust model, {\tt DustEM}.
\end{abstract}

\section{Introduction}

Dust is a fundamentally important component of the Universe. The presence of sub-micron size particles in the interstellar medium (ISM) was definitely evidenced by \citet{trumpler30} through the reddening of starlight. These particles were first considered as a nuisance for the estimate of stellar distances. Since that time and through the study of the ISM, there were growing evidences that dust is an essential component in the life of the ISM and thus, for the evolution of galaxies.
The ubiquity of dust in any phase of the ISM (and in other galaxies) was truly revealed by the first infrared surveys, especially the Infrared Astronomical Satellite (IRAS).
Dust is now widely used to trace the ISM physical conditions \citep[e.g.][]{bernard2010} and structure \citep[e.g.][]{mamd2010}.
It has also been detected in the early universe, in quasars at redshift $z\,\ga6$ \citep[e.g.][]{beelen2006}.  

Despite the fact that it represents a small mass fraction of the ISM ($\sim$1\%) dust plays a crucial role for its physics and chemistry.  Among the most important impacts of dust onto the gas phase, we have (i) the photoelectric effect that is the main heating process of the diffuse gas in the ISM, (ii) the catalysis of H$_2$ formation on the dust surface, otherwise very inefficient in gas phase under the ISM conditions because it involves a three-body reaction and (iii) the screening of photo-dissociating photons that allows for the survival of molecules in the ISM.  
Regarding the role of dust for the thermal balance, it does not only contribute by heating the gas through photoelectric effect: 30\% or more of the UV-visible photons emitted by stars are absorbed by the dust that re-emits the absorbed energy in form of an infrared-millimeter (IR-mm) thermal radiation for which ISM is transparent. 
For the same reason, dust shapes the way a galaxy looks like at both UV-visible and IR-mm wavelengths.
Dust couples the magnetic field to the gas phase in dense low ionized medium while it is responsible for such low ionization state of the gas through radiation field screening. Consequently, it plays an essential role in the last steps of core collapsing during stellar formation.  Dust can also dynamically couple the radiation field to the gas since it is more subject to radiation pressure and can drag the gas.

\begin{figure}
   \centering
      \includegraphics[width=1.\textwidth, angle=0]{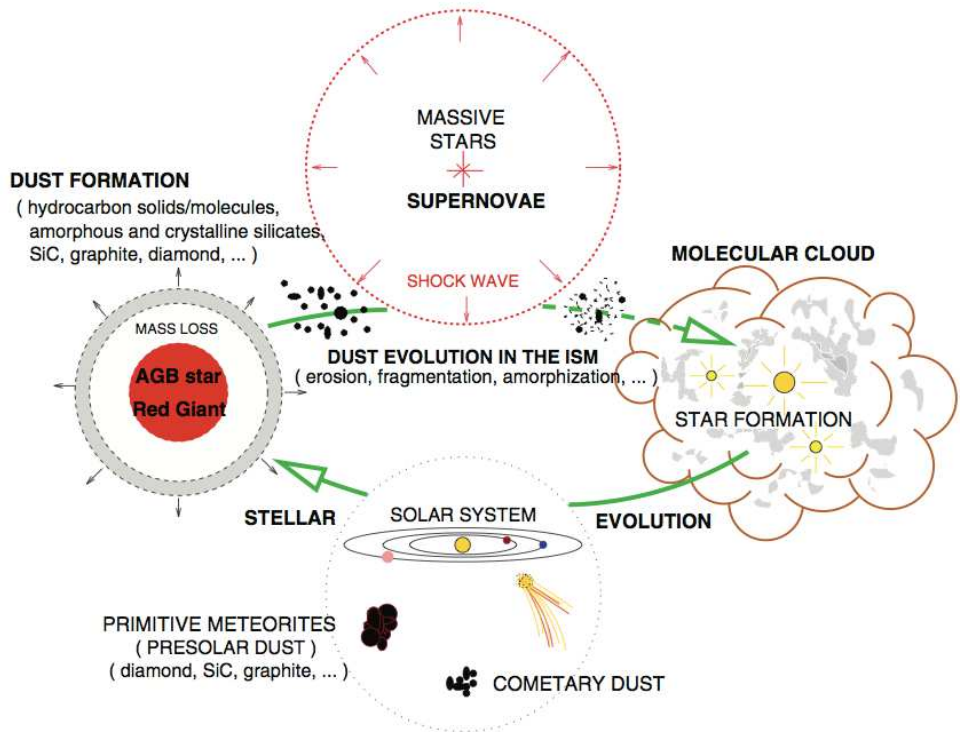}
     \caption{Schematic view of the dust evolution cycle driven by
       stellar\,$\leftrightarrow$\,ISM matter cycle \citep[from][]{jones2004}.
       Note that the dust lifecycle also comprises other shortcuts and sub-loops (e.g. diffuse\,$\leftrightarrow$\,dense ISM cycling, see \S\,\ref{sect:dustev}).
                      }
          \label{fig:dustev}
\end{figure}

It is now well established that dust properties evolve depending on the physical conditions of the ISM. 
Fig.\,\ref{fig:dustev} shows a schematic view of the dust evolution driven by the stellar\,$\leftrightarrow$\,ISM matter cycle.
There is an interplay between the dust and the rest of the ISM since the evolution of the ISM physical properties 
cause an evolution of the dust properties and subsequently an evolution of the dust impact on the ISM. 
It is necessary to characterize the physical processes responsible for the dust evolution to be able to quantify the dust impact 
throughout the ISM lifecycle.
Assuming some dust properties, the dust emission is widely used to estimate crucial quantities like cloud masses or star forming activity. However, to derive properly these quantities, one would need to take into account dust properties evolution, how it affects its emission and how it reflects the ISM properties. Here again, to characterize the dust evolution processes is of prime importance.

Dust study is a very broad topic including astronomical observations from X-ray to millimeter range as well as theoretical and laboratory study of the physics and the chemistry of ISM dust analogs (from large molecules to bulk solid). The present paper does not intend to give an exhaustive review about dust.
Moreover, it focuses on interstellar dust that although being evolutionary linked to circumstellar and protoplanetary dust, have quite different properties \citep[see for exemple table 2 of][]{jones2009}.
For more information, the reader can refer to numerous other reviews \citep[e.g.][]{boulanger2000a, Draine2003b, draine2004}, books and dedicated conference proceedings \citep[e.g.][]{Whittet2003, kruegel2003, CosmicDust, DustLesHouches2009}. 
An historical view of dust study can be found in \citet{li2003} and \citet{Li2005}.

In this paper, we first list the dust observables in section\,\ref{sect:observables} and further detail the depletion (\S\,\ref{sect:depletion}), extinction (\S\,\ref{sect:extinction}) and thermal emission (\S\,\ref{sect:emission}).
In section\,\ref{sect:dustev}, we discuss the dust lifecycle and evolution.
We address dust modeling, focusing on the {\tt DustEM} model in section\,\ref{sect:dustem}.
Concluding remarks are given in section\,\ref{sect:conclusion}.

\section{Dust observables}\label{sect:observables}

There are numerous direct and indirect observables that can bring constraints on interstellar dust properties.
\citet{dwek2005} gave a list of theses observables :
\begin{itemize}
\item the extinction, obscuration, and reddening of starlight;
\item the IR emission from circumstellar shells and different parts of the ISM (diffuse H I, H II regions, PDRs, and molecular clouds, X-ray emitting plasma);
\item the elemental depletion pattern and interstellar abundances;
\item the extended red emission (ERE) seen in various nebulae;
\item the presence of X-ray, UV, and visual halos around time-variable sources (X-ray binaries, novae, and supernovae);
\item the presence of fine structure in the X-ray absorption edges in the spectra of X-ray sources;
\item the reflection and polarization of starlight;
\item the microwave emission, presumably from spinning dust;
\item the presence of interstellar dust and isotopic anomalies in meteorites and the solar system; and
\item the production of photoelectrons required to heat neutral photodissociation regions (PDRs).
\end{itemize}
Even if related to the emission/extinction item, we can add to that list (i) the diffuse interstellar bands (DIBs) seen in the optical/near-IR extinction
that was first detected 90 years ago and (ii) the blue luminescence. 
The DIB carriers identification is one of the most challenging problem in the field of dust study since about 400 such
bands are observed while none of it was identified yet. It demonstrates the great importance of laboratory measurements to 
characterize more and more ISM dust candidates (in that case, very large molecules). 

These observational constraints span a broad range of environments and then various dust properties. Among these observables, the extinction curve, the IR-mm thermal emission and the depletion are the most often used to draw a picture of the average dust properties and how they evolve regarding the ISM properties.  We thus describe these observables in the following sections.

\section{Depletion} \label{sect:depletion}

\begin{table}[t]
    \caption{The elemental abundances (see \S\,\ref{sect:depletion} for references) in unit of $\rm{[X/10^6H]}$ (or ppm).}
  \centering
  \begin{tabular}{l l c c c c c}   
 \hline                 
                      &    &   C   &   O     &  Mg   &    Si    &   Fe    \\
 \hline
  Total &  Sun      & 269$\pm$33 & 490$\pm$60 & 40$\pm$4 & 32$\pm$2 & 32$\pm$3 \\
 &  F,G stars    & 358$\pm$82 &445$\pm$156  & 42.7$\pm$17.2 & 39.9$\pm$13.1 & 27.9$\pm$7.7\\
\hline
 Gas    &     & 75$\pm$25 & 319$\pm$14 & $\sim$0 & $\sim$0  & $\sim$0 \\
\hline
  Dust &  Sun & 194$\pm$41 & 171$\pm$62 & 40$\pm$4 & 32$\pm$2 & 32$\pm$3 \\
 & F,G stars & 283$\pm$86 & 126$\pm$157  & 42.7$\pm$17.2 & 39.9$\pm$13.1 & 27.9$\pm$7.7\\
  \hline
 \end{tabular}
   \label{tab:abundances}   
 \end{table}

In the ISM, most of the elements heavier than He are at least partially depleted in solid phase.  
As a consequence, the dust to gas mass ratio is correlated to the metallicity \citep[see][for nearby galaxies]{draine2007a}.
The solid phase abundance of an element can be estimated by comparing its abundance in gas phase with respect to its overall abundance in the ISM that is estimated by measuring its abundance in stellar atmospheres where no solid matter can subsist. 
Following \citet{jones2000}, we can separate the dust elemental constituents in four categories regarding their abundance in solid phase: 
(i) the primary elements, C and O, (ii) the secondary elements, Mg, Si and Fe, (iii) the minority elements, Na, Al, Ca and Ni and (iv) the trace elements, K, Ti, Cr, Mn, Co.  

Most of the current dust models only take into account the primary and secondary elements.
Table\,\ref{tab:abundances} lists the measured abundance and the inferred dust abundance for these elements.
The solar abundances are from \citet{asplund2009} while the F, G stars abundances are from \citet{sofia2001}.
The gas phase abundances are from \citet{dwek97} for the carbon and from \citet{meyer98} for the oxygen.
The \citet{dwek97} value for the carbon gas phase abundance was lower than some other measurements \citep[e.g. 140$\pm$20\,ppm from][]{cardelli96}
but is in good agreement with more recent estimates by \citet{sofia2009}.  
The gas abundances were measured toward diffuse lines of sight so that these abundances rely to refractory dust material (no ices, see \S\,\ref{sect:extinction}).

One should note that there might not be any stellar standard that well represents the ISM overall abundances because the processes of sedimentation and/or ambipolar diffusion during stellar formation could lower the heavy elements abundance in stars \citep{snow2000}. That could especially be true for B stars that were used as a probe of interstellar abundances since their composition was thought to be more representative of the current ISM one. Indeed, B stars have a lower metallicity regarding the Sun and young ($\le2$\,Gyrs) F, G stars
that seem to be more representative of the ISM metallicity.
Finally, we emphasis that \citet{draine2009a} argued that regarding the current uncertainties on elemental abundances and the various assumptions that are made for the dust 
properties, a model that departs from the measured elemental solid phase abundances by tens of percent should still be considered as viable.

\section{Extinction}\label{sect:extinction}

\begin{figure}
   \centering
      \includegraphics[width=0.75\textwidth, angle=0]{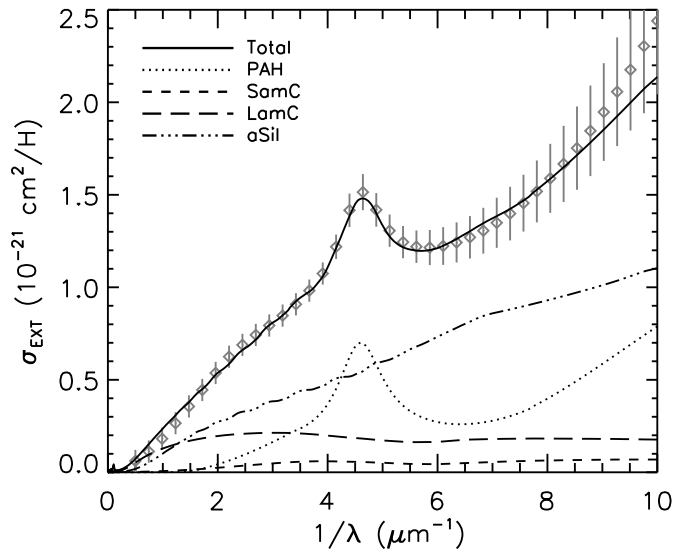}
   \caption{The extinction curve for dust in the diffuse ISM. The grey diamonds show the \citet{fitzpatrick99} extinction law for $\rm{R_V}\,=\,3.1$
and $N_H/E(B\,-\,V)\,=\,5.8\times10^{21}$ H cm$^{-2}$ \citep{bohlin78}. 
The black lines are the {\tt DustEM} model (see \S\,\ref{sect:dustem}).
                       }
          \label{fig:extuv}
\end{figure}

Fig.\,\ref{fig:extuv} shows the averaged extinction curve for the diffuse ISM.
Various spectral features allow us to infer the composition and mineralogy of dust.  
The shape of the extinction curve and its level normalized to the gas mass provide an information 
on the size distribution and abundances of the dust \citep[see][]{Weingartner2001}. 

The strongest observed dust feature is seen at 2175\,\AA\, and has a remarkably constant position while its width varies
from a line of sight to an other. 
The  $\pi \rightarrow \pi^{\star}$ electronic excitation of aromatic carbon (sp$^2$ hybridization) provides the most
straightforward explanation for this feature. However, the exact nature of such carbonaceous material remains unclear.
About 20-30\% of the carbon cosmic abundance is required to reproduce the strength of this feature.  
The $\sigma \rightarrow \sigma^{\star}$ electronic excitation of aromatic carbon produces an another bump at 750\,\AA\, 
whose red wing provides a natural explanation for the non-linear FUV rise ($\rm{6\,\la\,1/\lambda\,\la\,10\,\mu m^{-1}}$).

Narrower features at 3.3 and 3.4\,$\mu$m are attributed to C-H stretching mode in aliphatic (chain like) and aromatic hydrocarbons, respectively.
By comparing their observed profile with laboratory measurements, \citet{pendleton2002} concluded that the hydrocarbons in the diffuse ISM are
$\sim$15\% aliphatic and $\sim$85\% aromatic \citep[see also][]{dartois2007a}.

Broad bands are seen at 9.7 and 18\,$\mu$m that are attributed to Si-O stretching and O-Si-O bending mode in amorphous silicates, respectively.  
These bands are strong and require that most of the available Mg, Si, and Fe (if entering the composition of interstellar silicates) are locked in silicates.
The exact composition of such silicates is unknown.  
The absence of fine structures within these broad bands gives a higher limit to the crystallinity.
\citet{kemper2005} gave a limit of $\sim$2\% while more recently \citet{li2007} gave a limit of $\sim$3-5\,\%
taking into account the presence of ice mantle on silicate grains.

Indeed, infrared spectroscopy reveals the presence of extinction features
toward dense molecular material (A$_V\,\ga\,3$ on the line of sight, ${\rm n_H\ga10^3 - 10^4\,H\,cm^{-3}}$)
attributed to ices, mainly H$_2$O, CO$_2$ and CO. For a review, see \citet{dartois2005a}.

Starlight dimmed by dust appears partially polarized. 
Some grains must then be elongated and aligned with the magnetic field.
The decrease of polarization fraction from the visible to the UV \citep{serkowski73} implies
that only the bigger grains are efficiently aligned while the smaller grains are not aligned and/or not elongated. 
On the other hand, the 9.7\,$\mu$m silicate band is polarized while only a higher limit can be estimated for
the 3.4\,$\mu$m hydrocarbons band on the same line of sight \citep[e.g.][]{mason2007}.
Although such observations should be pursued, this result may indicate that big carbonaceous (not aligned) and 
silicate particles (aligned) are at least partially physically separated populations.

In the visible range, the albedo of $\sim$\,0.4\,-\,0.6 and the extinction continuum both require the presence of relatively big grains (${\rm a\sim0.1\,\mu m}$).
The high albedo is most likely produced by silicate material that are quite ``white'' regarding hydrocarbons.
The strong increase of extinction from the visible to the vacuum ultraviolet can not be reproduced without a significant amount of very small particles (a\,$\la$\,10\,nm). 
The shape of the extinction curve for a grain with a\,$\ga$\,10\,nm depends on its size while it is independent of the size
for a smaller grain (in Rayleigh regime). Consequently, only the abundance of the smaller grains is constrained by the extinction curve, not its size distribution, 
particularly its lower limit.

\section{Emission}\label{sect:emission}

Most of the energy absorbed by dust is radiated as IR-mm thermal emission that we describe here.
Although most often representing a negligible fraction, heating mechanisms other than the absorption of photons
can be efficient in some specific cases like gas-grain collisions in very hot gas of a supernovae (SN) remnant\footnote{Note that the very hot gas appellation does not apply for T\,$\sim10^4\,$K like in HII regions where the heating by UV photons dominates anyway.} \citep[see][]{dwek92}. 
Unfortunately, we do not have space in this review to discuss the blue luminescence, the ERE and the near-IR ($\rm{\lambda\sim\,1\,\mu m}$) continuum emission \citep[e.g.][]{wada2009, duley2009, flagey2006} nor the millimeter non-thermal emission of fast spinning grains \citep[e.g.][]{ysard2010, ysard2010a}.

\begin{figure}
    \centering
      \includegraphics[width=1.\textwidth, angle=0]{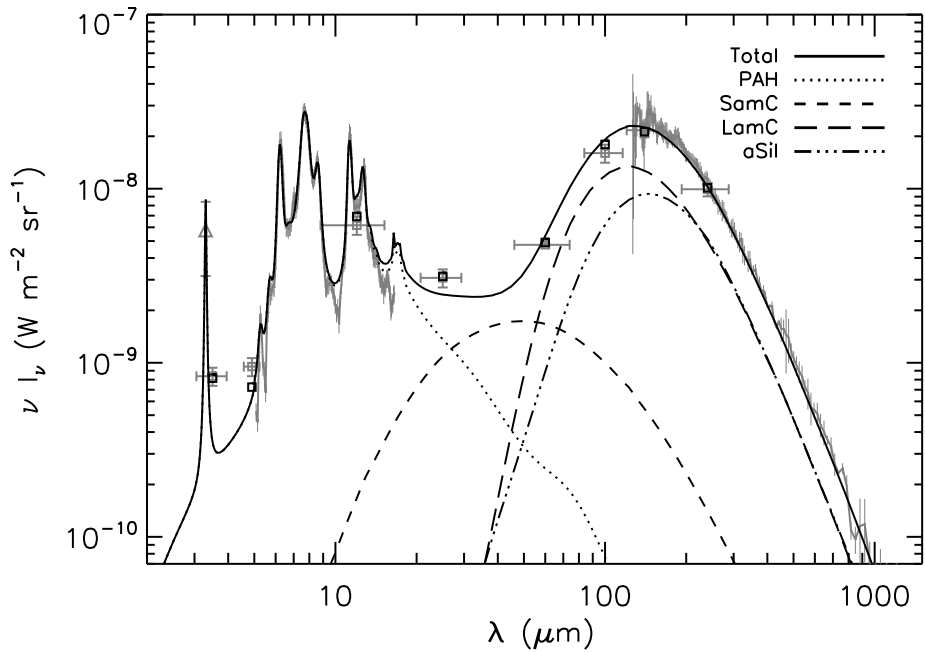}
      \caption{Dust emission for the DHGL medium.  
        Grey symbols and curves indicate the observed emission spectrum (AROME, ISOCAM/CVF, DIRBE, FIRAS) for $\rm{N_H\,=\,10^{20}\,H\,cm^{-2}}$
       \citep[see][ for details]{compiegne2010}.  Black lines are the 
       {\tt DustEM} model output (see \S\,\ref{sect:dustem})
         and black squares the modeled DIRBE points.
                      }
          \label{fig:dustem_emiss}
\end{figure}

Fig.\,\ref{fig:dustem_emiss} shows the averaged dust spectral energy distribution (SED) for the Diffuse High Galactic Latitude (DHGL) medium ($\rm{|b|>15^\circ}$)
obtained by \citet{compiegne2010} by correlating the IR-mm data with HI-21\,cm data.

The bigger grains (BG, $\rm{a>10\,nm}$) are in thermal equilibrium and emit a grey body spectrum in the far-IR. 
One has to keep in mind that the observed spectrum results from the emission 
of dust distributed in size and with various compositions that are accordingly distributed in temperature.
A physical dust model (see \S\,\ref{sect:dustem}) is then required for a detailed study. 
However, the far-IR spectrum is often analyzed by fitting a modified blackbody (i.e. grey body) to derive an effective temperature, $\rm{T_{eff}}$, and a
spectral index $\beta$ for the emissivity $\rm{\sigma_{em}(\lambda)\,=\,\sigma_{abs}(\lambda) \propto \lambda^{-\beta} }$ \citep[e.g.][]{dupac2001, dicker2009, bernard2010}.
The values derived from DIRBE and FIRAS for the diffuse interstellar medium are $\rm{T_{eff}\,=\,17.5\,K}$ 
and $\beta\,=\,2$ \citep[see][]{boulanger96}. 
Taking U to be a scaling factor with respect to the average interstellar exciting radiation field \citep[e.g. the one of][]{mathis83}, we
can estimate $\rm{T_{eff}}$ with the formula $\rm{T_{eff}\,=\,17.5\,U^{1/(4+\beta)}\,K}$.  
Note that the observed $\beta$ also varies significantly over the sky \citep[e.g.][]{dupac2003, desert2008} possibly resulting from dust properties evolution and some microphysical effects in amorphous dust grains like the two levels system effect and the disordered charge distribution effect \citep[see][]{meny2007}. 
Although they enclose most of the dust mass, the properties of the BGs remain poorly known. 
New observations currently carried out with Herschel and Planck satellites will bring a new insight on those properties.

\begin{figure}
    \centering
      \includegraphics[width=.9\textwidth, angle=0]{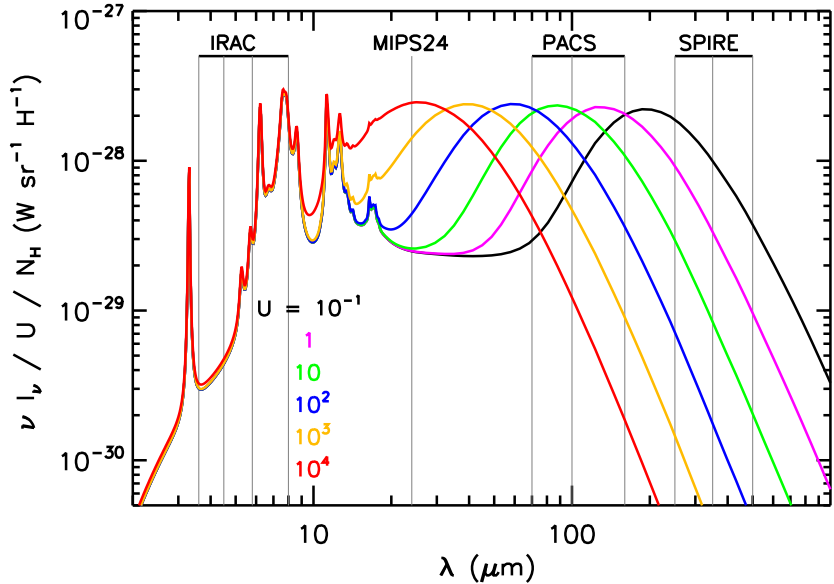}
      \caption{The modeled dust SED per H atom for different scaling factors of the \citet{mathis83} exciting radiation field, U=0.1 to 10$^4$.
                    All spectra have been divided by U to emphasize changes of the SED shape.
                    Also shown are the photometric band positions of Herschel and Spitzer satellites.
                    (Color version online)
                     }
          \label{fig:SEDvsU}
\end{figure}

The IRAS observations revealed the ubiquity of dust mid-IR emission.
Such a short wavelength emission requires high temperatures ($\rm{T_{eff}\ga100\,K}$)
that can not be reached by classical BG ($\rm{a\,>\,10\,nm}$) at thermal equilibrium unless $\rm{U\ga10^4}$.
However, a very small grain ($\rm{a\sim1\,nm}$) that has a small heat capacity regarding the energy carried by a single UV-visible photon can
reach high temperature under single photon absorption event. 
Following such an absorption, it cools down to the fundamental level until the next absorption. 
For the total spectrum emitted by dust to be computed by accounting for such a stochastic heating,
one has to derive the probability distribution, dP/dT,
of finding a given grain at a temperature in the range [T, T+dT] \citep[see][]{Desert86} and then to integrate
over T, over all sizes and over all grain types
the emitted spectra weighted by dP/dT.
The normalized dP/dT of a stochastically heated particle does not depend on the intensity
of the exciting radiation field (i.e. photon absorption rate) as long as the particle undergo single photon events (i.e. cools down to the fundamental level before a new absorption).
It rather depends on the average energy of the absorbed photons.
Consequently, unlike the grains at thermal equilibrium, the spectral shape of the smallest particles emission
does not depend on the intensity of the exciting radiation field (U) as seen on Fig.\,\ref{fig:SEDvsU}.
Note that for a grain at thermal equilibrium, dP/dT tends toward a Dirac function, $\rm{\delta(T-T_{eq})}$.

As a result of the stochastic heating, the shape of the near- and mid-IR SED is very sensitive to 
the size (i.e. the heat capacity) of the very small 
particles and brings a constraint on its distribution, especially its lower limit.
The presence of very small particles ($\rm{a\la1\,nm}$) is required to explain the observed emission down to $\rm{\sim3\,\mu m}$
corresponding to $\rm{T\sim1000\,K}$.
On the other hand, theoretical works have shown that polycyclic aromatic hydrocarbon
particles (the smallest ISM dust) smaller than $\sim$3.5$\AA$ ($\sim$20 carbon atoms) could not survive in the diffuse ISM due
to photolysis \citep[e.g.][]{allamandola89}.

Strong emission features at 3.3, 6.2, 7.7, 11.3 and 12.7\,$\mu$m (see Fig.\,\ref{fig:dustem_emiss}) were first observed in NGC7027 by \citet{gillett73}.
Using the Infrared Space Observatory and Spitzer satellites, these features (plus other weaker ones) were detected toward almost every lines of sight where dust is present, 
spanning a broad range of physical conditions.
They were first called unidentified infrared bands (UIBs) and were later attributed to the IR fluorescence
of UV pumped Polycyclic Aromatic Hydrocarbons \citep[PAH,][]{duley81, Leger84, allamandola85}.
Although no specific terrestrial PAH analog was identified as the carrier, 
such a form of carbonaceous material is likely to produce these bands that are now called PAH features.
Interstellar PAHs must enclose $\sim$20\% of the overall carbon abundance and must consequently
account for a significant fraction of the 2175\AA\, bump of the extinction curve (see \S\,\ref{sect:extinction}).
Due to this large abundance and small size, PAHs must dominate the dust surface and then be a major contributor for
the photoelectric heating and for surface reaction (e.g. H$_2$ formation).
On the other hand, the bright PAH features spectrum can be used to trace the ISM physical conditions 
and as a probe of star formation activity in the local and distant Universe.
PAH has accordingly become an important topic of which a recent review is given by \citet{tielens2008}.

\section{The dust lifecycle \& evolution}\label{sect:dustev}

Dust formation has been evidenced in outflows of AGB stars and in planetary nebula.
Carbon rich stars mainly produce carbonaceous dust (in some cases SiC) while oxygen
rich stars mainly produce silicates.
These silicates have a crystallinity up to $15\%$ \citep{waters2004} which is 
higher than for ISM silicates (see \S\,\ref{sect:extinction}). Silicate stardust after being injected into the ISM
must then be amorphized, possibly by cosmic ray bombardment \citep[e.g.][]{demyk2004}. 
Note that although they can alter it, cosmic rays are not an agent of dust destruction.
Dust formation is also observed in supernova ejecta. However, the amount of dust formed in SN that is actually
injected into the ISM is controversial since dust destruction also occurs in SN remnants.

Once they are injected into the ISM, dusts are subjected to many processes that are likely to modify their
properties: (i) thermal sputtering by the gas in high-velocity ($\rm{>200\,km\,s^{-1}}$) shocks; (ii) vaporization and shattering by grain-grain collisions in lower velocity shocks; (iii) alteration/photolysis under cosmic ray or hard photon absorption; and (iv) accretion and coagulation in dense molecular clouds. 
A description of most of these processes and grain lifetimes in the ISM was presented by \citet{jones2004}.

The dust destruction mainly takes place in the diffuse medium 
under SN shock wave occurrence. 
The destruction timescale 
was estimated to be $\rm{\tau_{destruction}\,\sim\,5\,10^{8}\,yrs}$
while considering all possible sources, the dust formation timescale was estimated to be $\rm{\tau_{formation}\,\sim\,3\,10^{9}\,yrs}$ \citep[e.g.][]{jones94}. 
A factor of almost 10 between these timescales is inconsistent
with the high abundance of dust observed in the ISM.
Dust must then be efficiently regenerated in the ISM itself.

Dust has been detected in early Universe objects (quasars, luminous galaxies at z$\sim$6).
The AGB stars whose progenitors are $\rm{M_\star\sim1.5-3\,M_\odot}$ dominate the dust injection in the disk of mature galaxies.
These stars reach the giant branch only after $\sim$700\,$\rm{Myrs}$ and could then not be responsible for the 
dust production in the early Universe.
On the other hand, SN in order to explain the dust mass measured in the early Universe
should produce $\rm{M_{dust}\ga1\,M_\odot}$/SN \citep{dwek2007}, a rate that exceed the
observed rate in the local Universe ($\rm{M_{dust}\la0.02\,M_\odot}$/SN).  
This also leads to the conclusion that dust mass must significantly increase in the ISM itself.

The physical conditions in the ISM are not suitable for dust nucleation.
However, in the dense ISM (shielded from UV radiation field and SN shock waves), dust can grow by coagulation and by accretion of gas.
The timescale for dust ``regeneration'' by accretion in a dense cloud ($\rm{n_H\sim10^4\,H\,cm^{-3}}$) is indeed smaller than the cloud life time 
which is a few million years. 
Note that the coagulation (of dust grains together) modifies the size distribution but does not increase the total dust mass.

The observed ISM dusts must have been partially destroyed and regenerated multiple times after their nucleation
and must have very different properties compare to ``stardusts'' \citep[e.g.][]{zhukovska2008}.
Particularly, although nucleated in different stellar environments, silicates and carbonaceous dusts should be mixed by subsequent	accretion and coagulation in the ISM. 
However, this evolutionary scenario seems contradictory with the fact that the bigger silicates and carbonaceous grains
are at least partially physically separated in the ISM (see \S\,\ref{sect:extinction}).
Although this inconsistency can partly be circumvented by invoking chemically selective accretion processes \citep[see the description by][]{draine2009b},
it would still hold for the produce of coagulation. 
This points out the need to know in further details the processes that drives the dust nucleation and its further
evolution and destruction in the ISM.

We finally note that a significant dust evolution in the ISM is supported by many observational evidences.
The evolution of the extinction curve \citep[e.g.][]{fitzpatrick90} mainly traces changes in the size distribution \citep[e.g.][]{kim94a}. 
The release (the accretion) of gas species from (onto) the dust can be reflected by variations of the depletion. Such variations
are particularly noticeable in SNR where the refractory elements return to the gas phase due to dust destruction \citep[e.g.][]{slavin2008}.
The SED also evolves \citep[e.g.][]{boulanger90, abergel2002, berne2007} tracing 
an evolution of the size distribution \citep[e.g.][]{mamd2002, stepnik2003, compiegne2008, flagey2009} or/and
of the physical properties of the emitters \citep[e.g.][]{berne2007, compiegne2007}.

\begin{figure}
    \centering
      \includegraphics[width=.7\textwidth, angle=0]{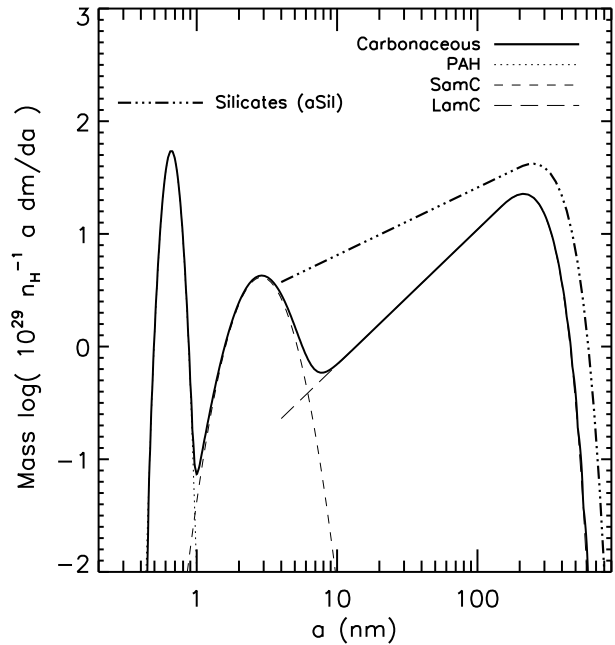}
      \caption{Mass (size) distribution for the dust components used in {\tt DustEM} model to reproduce the DHGL observables.
                    The population of amorphous carbon dust is split into small (SamC) and large (LamC) grains.
                     }
          \label{fig:sdist}
\end{figure}

\section{Dust modeling and the {\tt DustEM} model} \label{sect:dustem}

 Physical dust models are necessary tools to infer interstellar dust properties from the observations.
A dust model should ideally satisfy all observable constraints but also only make use of
dust components whose presence is plausible in view of its likely formation and destruction in the ISM.
The history of modern dust models has begun with \citet{MRN1977}.
Like this latter one, early models focused on explaining the extinction (some of them also describing polarization by extinction) using 
silicates and carbonaceous dust as the grain materials \citep[e.g.][]{draine84, kim95}. 
Some models suggested the presence of organic refractory mantle on the surface of the bigger grains, resulting 
from the accretion of an ice mantle ($\rm{H_2O}$, $\rm{CH_4}$, $\rm{NH_3}$) further processed by UV photons \citep[e.g.][]{greenberg95,li97}.
The \citet{desert90} model (hereafter DBP90) was the first that consistently reproduced both extinction and emission, dealing with the stochastic heating mechanism.
Most of the current models aim to consistently reproduce the observed extinction and emission \citep[][hereafter DL07]{siebenmorgen92, dwek97, li2001, zubko2004, draine2007} as well as the associated polarization \citep[][]{draine2009}.

Here we focus on the {\tt DustEM} dust model that is described in \citet{compiegne2010}.
Three dust components are used in this model : (i)\,PAH, (ii)\,hydrogenated amorphous carbon (HAC or amC) and (iii)\,amorphous silicate. 
PAH optical properties are empirical \citep[described in][and DL07]{li2001} 
while the optical efficiencies for the HAC and silicate grains are obtained from the refractive index 
of bulk material using Mie theory and assuming spherical particles (the simplest approximation).
The mass distribution shown on Fig.\,\ref{fig:sdist} allows for the reproduction of the observed SED 
(Fig.\,\ref{fig:dustem_emiss}), extinction curve (Fig.\,\ref{fig:extuv}) and depletion of the DHGL medium. 
The observables related to the DHGL medium are commonly used to constrain dust properties and provide a natural reference 
to compare with when studying dust evolution.

Note that interstellar grains might include components other than hydrocarbons and silicates,
for example silicon carbide (SiC), metal oxides (e.g. MgO, FeO), metallic form of Fe and the minority and trace elements listed 
\S\,\ref{sect:depletion}. However, the abundance of such components is 
sufficiently low to be neglected in models with the current level of sophistication \citep[see also \S\,2.4 of][]{draine2009a}.

Carbonaceous dust (other than PAH) is often considered to be in the form of crystalline graphite because it can explain the 2175$\AA$ extinction band,
but PAH can also account for this band.
Moreover, graphite cannot explain the band profile variability observed in the ISM \citep{fitzpatrick2007,Draine2003b}. 
On the other hand, throughout the dust lifecycle, from dense molecular cores to the diffuse ISM, the grain populations are the result of complex, non-equilibrium evolutionary processes and it appears natural to consider that carbon dust is amorphous, as observations have clearly demonstrated for silicate grains.  
Finally, models show that graphite particles are not efficiently destroyed when encountering a fast shock ($\rm{V_S\sim100-150\,km\,s^{-1}}$) while
80 - 100\% of carbon atoms are actually observed to be in the gas phase under such conditions.
Conversely, HAC particles seem to be efficiently destroyed.
These are the reasons why HAC is used as the carbonaceous grain population
in the {\tt DustEM} DHGL model \citep[see][for the full discussion]{compiegne2010}.

{\tt DustEM} comprises a SED fitting tool and also allows for the empirical variation of the long wavelength emissivity of dust.
Although not presented in \citet{compiegne2010}, the model
deals with the spinning dust millimeter emission and polarization (in extinction and emission).
The source code of this model is written in 
{\sc fortran90} and is available online \citep[see][]{compiegne2010}.

\section{Conclusion}\label{sect:conclusion}

Although the broad lines of dust properties and 
lifecycle are known, we still miss a detailed understanding since
not even for the principal dust constituents 
(silicate and carbonaceous materials) the exact properties are clear.
Furthermore, the carriers of some observables like the DIBs, the ERE, the blue luminescence and the near-IR
continuum are poorly or even not identified and are then not already
included in the general picture of dust models.
Details of the efficiency of dust formation, its injection into the ISM, its further evolution and destruction need to 
be further investigated.
Among the current observing facilities, Herschel and Planck, when combined to
Spitzer and IRAS, give access to the entire dust SED (see Fig.\,\ref{fig:SEDvsU}) and 
allow us to follow its evolution over a large fraction of the sky (i.e. over a broad range of 
physical conditions) at scales down to tens of arcseconds.
These observations, analyzed with a physical dust model like {\tt DustEM} \citep[see][]{abergel2010, compiegne2010a}
will certainly provide a new insight on dust properties, its evolution and its role in the ISM.

\acknowledgements I wish to thanks the organizers of this conference, especially Roland Kothes and Tom Landecker, for the great 
job they have done and for the opportunity they offered me to give this review talk.

\bibliography{compiegne_mathieu}

\end{document}